\documentclass[preprint2, times]{aastex6}
\usepackage{amsmath}
\usepackage{mathtools}
\usepackage{microtype}
\usepackage[all]{hypcap}

\def\mchirp{\mathcal{M}}

\begin{document}

\title{Harmonic space analysis of pulsar timing array redshift maps}

\author{Elinore Roebber\altaffilmark{1}\hyperlink{email}{$^{\star}$} \& Gilbert Holder\altaffilmark{1,2}}
\affil{\altaffilmark{1}Department of Physics and McGill Space Institute, McGill University, 
			     3600 rue University,  Montr\'eal, QC, H3A 2T8, Canada \\
	\altaffilmark{2}Department of Physics, University of Illinois at Urbana-Champaign, 
				1110 W Green St Urbana, IL, 61801, USA 
	}
\email[\hypertarget{email}{$^\star$ }]{roebbere@physics.mcgill.ca}

\shorttitle{Harmonic space analysis of PTA redshift maps}
\shortauthors{Roebber \& Holder}

\begin{abstract}
In this paper, we propose a new framework for treating the angular information in the pulsar 
timing array response to a gravitational wave background based on standard cosmic microwave 
background techniques.  We calculate the angular power spectrum of the all-sky gravitational 
redshift pattern induced at the earth for both a single bright source of gravitational radiation and 
a statistically isotropic, unpolarized Gaussian random gravitational wave background. 
The angular power spectrum is the harmonic transform of  the Hellings \& Downs curve.  
We use the power spectrum to examine the expected variance in the Hellings \& Downs curve 
in both cases.  Finally, we discuss the extent to which pulsar timing arrays are sensitive to the 
angular power spectrum and find that the power spectrum sensitivity is dominated by the quadrupole
anisotropy of the gravitational redshift map.
\end{abstract}

\maketitle

\section{Introduction}

Pulsar timing arrays (hereafter \textsc{pta}s) are galactic-scale gravitational wave detectors 
based on the precise timing of millisecond pulsars across the sky \citep{foster90}.  The nanohertz 
frequency band of gravitational waves (\textsc{gw}s) accessible to \textsc{pta}s has several 
potential production mechanisms, the most prominent of which is due to the inspiral of subparsec 
supermassive binary black holes \citep[\textsc{smbbh}s; see][and references therein]{lommen15}.  

\textsc{Smbbh}s with chirp mass $\mchirp > 10^8 M_\odot$ at redshifts $z\lesssim 2$ are expected 
to produce most of the signal \citep[e.g.][]{sesana08}.  
Since there should be many such sources evolving over times much 
longer than human timescales, the \textsc{gw} signal is expected to form a stochastic background
with considerable source confusion.
However, individual strong sources may stand out \citep{sesana08,ravi12a}ß.

A passing \textsc{gw} induces compression and rarefaction of spacetime along its polarization axes.  
Periodic signals such as rays of light or pulse trains
propagating through this region will be blue- or redshifted according to the strain 
of the \textsc{gw}.  For periodic signals with frequency much higher than that of the \textsc{gw}, 
the shift will build up, producing a potentially measurable effect.  This is the principle on which 
several models of \textsc{gw} detection are founded, including interferometers such as \textsc{ligo}
\citep{ligo16b} and \textsc{lisa} \citep{elisa13} as well as for \textsc{pta}s \citep{lommen15}.  
There are three \textsc{pta} consortia: \textsc{epta} \citep{lentati15}, \textsc{nanog}rav \citep{arzoumanian16}, 
and \textsc{ppta} \citep{shannon15}.  They combine together to form the \textsc{ipta} \citep{verbiest16}.  

\textsc{Pta}s search for integrated red- and blueshifts produced by gravitational waves passing the 
earth through the careful timing of a network of millisecond pulsars across the sky.  Each millisecond 
pulsar produces an extraordinarily regular train of high-frequency pulses.  If this pulse train is redshifted
by a \textsc{gw} with typical strain $\lesssim 10^{-14}$ \citep[e.g.][]{lommen15}, 
no effect will be immediately visible, but after the
passage of many pulses, a difference between the expected and actual time of arrival of pulses will 
become apparent.  This timing residual is the basic measurable quantity for a \textsc{pta}.  

A \textsc{gw} of a given polarization will induce red- and blueshifts according to the geometry 
set by the direction of propagation of the \textsc{gw} and the projection of its polarization axes onto 
the sky.  In order to sample this effect as fully as possible, \textsc{pta}s time many millisecond pulsars
across the sky and search for a correlation in their timing residuals which reflects the redshift pattern
induced by \textsc{gw}s.  

The expected form of this correlation is the Hellings \& Downs curve \citep{hellings83}, 
which was originally derived for a statistically isotropic unpolarized Gaussian random field of 
gravitational waves.  It also represents the expected correlation 
pattern for a single \textsc{smbbh} source of \textsc{gw}s \citep{cornish13}.  

However,  the gravitational wave background (\textsc{gwb}) expected to be produced by a 
population of inspiraling \textsc{smbbh}s will be neither completely dominated by a single source 
nor a completely stochastic Gaussian field.  In general, it should be somewhere in between 
\citep[e.g.][]{sesana08}.  

Although much work has made use of the assumption that a stochastic background would have 
Gaussian statistics, single sources should not be neglected in the \textsc{pta} search for \textsc{gw}s
\citep{rosado15a}.  This is because the distribution of \textsc{smbbh} sources is such that the rarest 
brightest sources dominate the signal in the \textsc{gwb} 
\citep{sesana08,kocsis11,ravi12a,cornish13,roebber16}. 

In light of this, it is of interest to search for angular information in the \textsc{gwb}.  \textsc{Pta}s 
can be likened to a collection of gravitational wave antennas: their angular resolution is limited
but not nonexistent.  This has been taken advantage of in the attempt to search for individual 
sources and hotspots \citep[e.g.][]{sesana10, corbin10, babak12, simon14}. 
Additionally, recent works have characterized the correlation patterns expected for statistically 
anisotropic backgrounds made up of a large number of sources \citep{mingarelli13,taylor13} as 
well as attempting to map general \textsc{gwb}s \citep{gair14,cornish14}.  

Many of these recent works have focused on estimating the distribution of gravitational wave 
signals produced by the source population, either in terms of power or components of the 
gravitational wave tensor.    
However, the gravitational wave strain is not directly measured by \textsc{pta}s.  The large 
effective beam patterns smear power out across the sky, mixing contributions from different 
sources.  Furthermore, since gravitational waves are tensors and the timing residuals 
measured by \textsc{pta}s are scalars, there are components of the strain that cannot be 
measured \citep{gair14}.  Both of these complications can be sidestepped by working with the 
maps of the theoretical timing residuals or equivalently, the redshifts induced by the passing 
gravitational waves.

In this paper, we consider an alternate analysis of the \textsc{gwb} in the \textsc{pta} band, 
inspired by standard cosmic microwave background (\textsc{cmb}) methods.  Our primary 
quantity of interest is the redshift induced in all directions on the sky by \textsc{gw}s passing 
the earth.  This is related to pulsar timing residuals in the following fashion: 
\begin{itemize}
	\item Sampling the redshift field in a direction $\hat p$ gives the amount by which 
		the pulse train of a pulsar at $\hat p$ is redshifted or blueshifted due to the influence
		of \textsc{gw}s passing the earth.
	\item Integrating the redshift at $\hat p$ gives the shift in the pulsar's timing residuals 
		due to \textsc{gw}s passing the earth (the `earth term').
		 Since we limit our discussion to circular and non-evolving \textsc{gw} sources, 
		 the integrals are trivial.
\end{itemize}

Furthermore, we initially analyze redshift maps in harmonic space, and transform back to real space
when considering the implications.  This approach may not be practical for experimental analysis 
and we present it primarily as an alternate framework for understanding the angular information in 
the gravitational wave background.  

In \autoref{sec:formalism} we review the standard mathematical formalism underlying \textsc{gw}s
produced by circular, slowly-inspiraling binary systems and their measurement by \textsc{pta}s 
and produce example maps of the redshift patterns produced by various \textsc{gwb}s.
In \autoref{sec:Cl_analysis} we present our harmonic-space analysis of redshift maps and 
specifically discuss two limiting cases: a single \textsc{gw} source and a statistically isotropic
Gaussian random \textsc{gwb}.  In \autoref{sec:variance_Cl_HD} we discuss the relation between 
the two-point function in real and harmonic space and present a case where the harmonic 
analysis provides insight into real-space quantities: how variance in the power spectrum affects 
the shape of the Hellings \& Downs curve.  In \autoref{sec:Cl_SN} we discuss the degree to which
the power spectrum is measurable in an ideal \textsc{pta}.  And finally, in \autoref{sec:discussion}
we present our conclusions and discuss future directions.

\section{Gravitational wave formalism}
\label{sec:formalism}

A gravitational wave is a transverse plane wave propagating as spatial perturbations in the metric.  
It has a spin-$2$ symmetry and two polarizations \citep{maggiore08}: 
\begin{equation}
h_{ij}(t, \hat k) = h^+(t) \, e^+_{ij} (\hat k)+ h^\times(t) \, e^\times_{ij}(\hat k), 
\label{eq:hab_def}
\end{equation}
where $h^+$ and $h^\times$ are the amplitudes of the two polarizations, $e^{+}_{ij}$ and 
$e^{\times}_{ij}$ are the polarization tensors, and 
 $\hat k$ is the direction of propagation of the wave.  Sub- and superscripts $i,j$ are 
written using the Einstein summation notation and denote the tensorial nature of gravitational waves.  

The geometry of an incoming gravitational wave can be 
written in terms of a radial vector in the direction of propagation of the gravitational wave, and two 
vectors perpendicular to it which define a basis for the polarization of the wave.  Our choice of 
conventions follows \citet{gair14}:
\begin{align}
\hat{k} &= \sin\theta\cos\phi \, \hat{x} + \sin\theta\sin\phi \, \hat{y} + \cos\theta \, \hat{z} \nonumber \\
\hat{l} & = \cos\theta\cos\phi \, \hat{x} + \cos\theta\sin\phi \, \hat{y} - \sin\theta \, \hat{z} \nonumber \\
\hat{m} & = -\sin\phi \, \hat{x} + \cos\phi \, \hat{y}.
\end{align}
If we consider $\hat k$ to be a radial vector along the axis of propagation, the location of the 
gravitational wave source is in the $-\hat k$ direction, or equivalently at the angle on the sky $(\pi-\theta, \phi + \pi)$.
The perpendicular vectors $\hat{l}$ and $\hat{m}$ are vectors in the $\hat\theta$ and $\hat\phi$ directions defining the plus and cross polarizations of the incoming gravitational wave:
\begin{align}
e^+_{ij}(\hat k) &= \hat l_i \hat l_j - \hat m_i \hat m_j \nonumber \\
e^\times_{ij}(\hat k) &= \hat l_i \hat m_j + \hat m_i \hat l_j.
\label{eq:e+ex}
\end{align}

\begin{figure*}
\begin{center}
\includegraphics{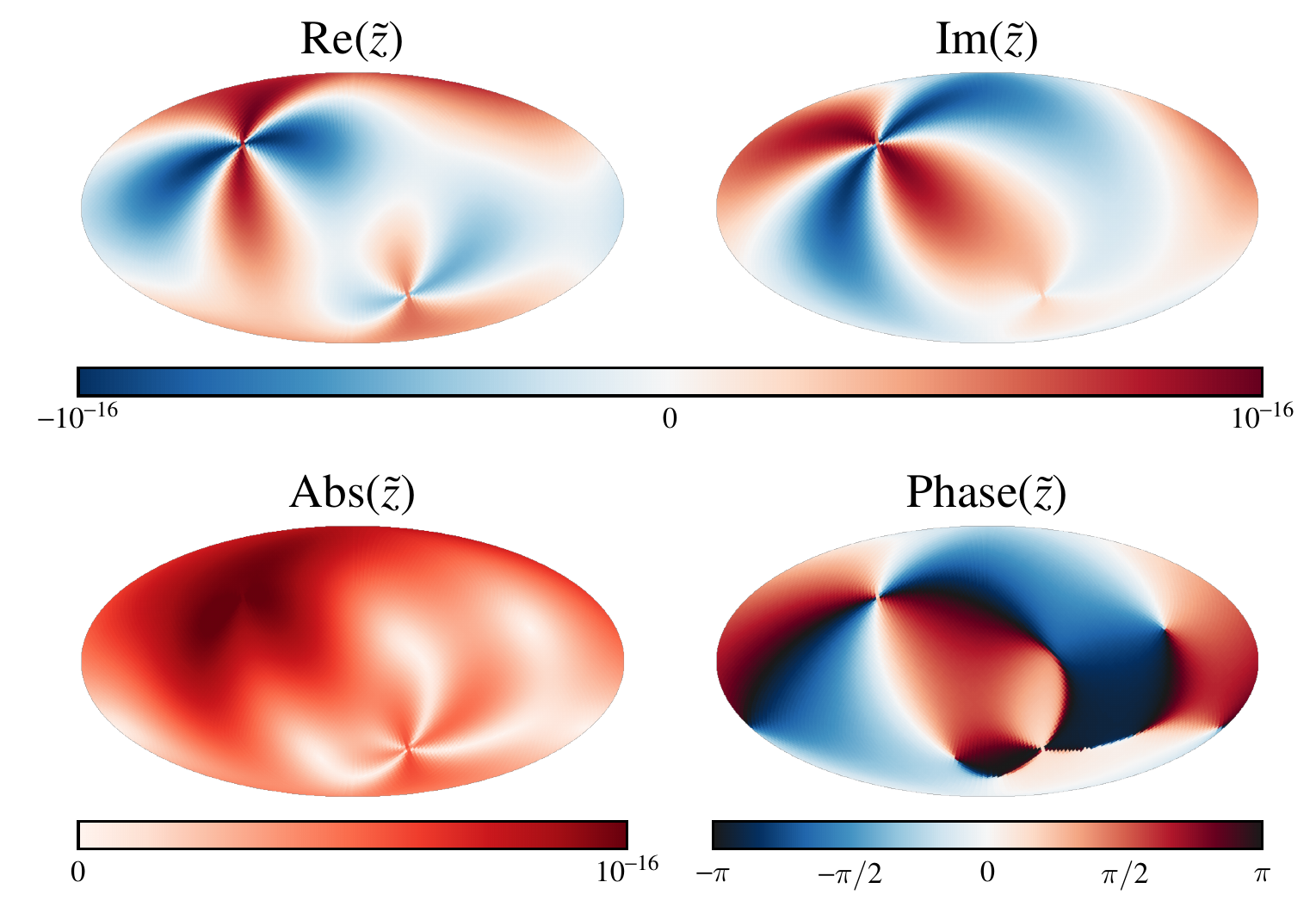}
\end{center}
\caption{Mollweide projection of two \textsc{gw} sources in the frequency domain with equal $A_\textsc{gw}$.  
		The source in the upper left is face-on and the 
		source in the lower right is edge-on.  Both have random initial phases and polarization angles.
		Face-on sources contain equal components in $+$ and $\times$ and have evenly distributed
		real and imaginary components. As a result, the amplitude of a face-on source is constant 
		in azimuthal angle.  In the time domain it rotates.
		By contrast, an edge-on source produces only $+$ polarization in its rest frame.
		It has a single redshift pattern split between the real and imaginary components and has stripes
		radiating out from its center which are neither redshifted nor blueshifted by the \textsc{gw}s.  
		In the time domain, it appears as a static redshift pattern which fades in and out as the binary 
		rotates.  It appears fainter than a face-on source since its $a(\iota)$ coefficient is smaller.
		Both kinds of sources show characteristic spin-2 phase patterns, in which points  separated by 
		a $90^\circ$ rotation around the source are out of phase.  The smoothly varying
		behavior and sharp edges of the two phase patterns reflect its rotation or lack thereof in the 
		time domain.}
\label{fig:two_srcs}
\end{figure*}

These are all general properties of \textsc{gw}s, but we are interested in \textsc{gw}s generated by 
\textsc{smbbh}s, which can be described more closely.  In particular, we restrict ourselves to the 
case where the binary is circular and very slowly evolving, so that we can ignore its evolution on 
observational timescales. Gravitational waves of this form can be described by four additional 
parameters: $(\mathcal{A}, \iota, \psi, \Phi_0)$, as described in the following paragraphs 
\citep[e.g.][]{cutler94,sesana10}.

The amplitude $\mathcal{A}$ contains information about the non-angular degrees of freedom of the 
binary. 
\begin{equation} 
\mathcal{A} = 2 \frac{(G\mchirp)^{5/3}}{c^4D}(\pi f_\text{emit})^{2/3},
\end{equation}
where $\mchirp = (m_1m_2)^{3/5}/(m_1+m_2)^{1/5}$ is the chirp mass of the binary system, 
$D$ is the proper distance, and $f_\text{emit}$ 
is the frequency of the gravitational wave in the binary's rest frame.  

The inclination $\iota$ of the binary tells us the relative contribution of each polarization.  A face-on 
or face-off binary is circularly polarized, and produces equal quantities of the plus and cross polarizations. 
An edge-on binary only produces plus polarization, and can be considered to be linearly polarized.  
A general binary is somewhere in-between, and its \textsc{gw} is elliptically polarized.
For an inclination $\iota$, the contributions to the plus ($a$) and cross ($b$) polarizations can be 
expressed as
\begin{align}
	a(\iota) &= 1 + \cos^2 \iota \nonumber \\
	b(\iota) & = -2 \cos \iota.
\end{align}

The angle $\psi$ encodes the transformation between GW polarizations 
between the source coordinate system and that of the observer.  It gives the degree to which 
the plane of the binary is misaligned with the $(\hat l, \hat m)$ basis given above, which leads to 
mixing between the different polarizations:  
\begin{align}
 h_+' &= h_+ \cos 2\psi + h_\times \sin 2\psi \nonumber \\
h_\times' & = -h_+ \sin 2\psi + h_\times \cos 2 \psi
\label{eq:hpsi_def}
\end{align} 
The mixing takes the form of a rotation by $2\psi$ since gravitational waves are spin-2: a pure 
$+$ mode becomes purely $\times$ if the $(\hat l, \hat m)$ coordinate system is rotated by 
$45^\circ$.  The angle $\psi$ and 
the angles giving the location of the \textsc{gw} source $(\theta, \phi)$  are defined in terms of 
the coordinate system of the observer, and can be changed by a rotation of the coordinate axes. 

The overall temporal phase of the binary is given by 
\begin{align}
	\Phi(t) = \int_0^t 2\pi f(t') dt' \approx 2\pi f t + \Phi_0,
	\label{eq:phase}
\end{align}
where $\Phi_0$ is the initial phase of the binary.  To make the approximation in \autoref{eq:phase}, 
we assume non-evolving circular binaries.

Altogether, the components of \textsc{gw}s produced by a non-evolving circular binary can be written:
\begin{align}
	h_+(t, \hat k) &= \mathcal{A} \left[ a\cos 2\psi \cos \Phi(t) + b\sin 2 \psi \sin \Phi(t) \right]  \nonumber \\
	h_\times(t, \hat k) &= \mathcal{A} \left[ b \cos 2\psi \sin \Phi(t) - a\sin 2 \psi \cos \Phi(t) \right]
\end{align}
Assuming that the binaries are circular and non-evolving, as in \autoref{eq:phase}, this may be 
easily be written in frequency space:
\begin{align}
	\tilde h_+(f, \hat k) &= \frac{\mathcal{A}}{2}   \big[ a \cos 2\psi (\cos \Phi_0 + i\sin\Phi_0) + \nonumber \\
			   & \quad \quad \,\,\,\,\, b \sin 2\psi (\sin \Phi_0 - i\cos\Phi_0) \big]
	\nonumber \\
	\tilde h_\times(f, \hat k) &= \frac{\mathcal{A}}{2}   \big[ b \cos 2\psi (\sin \Phi_0 - i\cos\Phi_0) - \nonumber\\
			   & \quad \quad \,\,\,\,\, a \sin 2\psi (\cos \Phi_0 + i\sin\Phi_0) \big],
	\label{eq:h(f)}
\end{align}
where $f$ is the positive frequency associated with the binary, and $\tilde h(f)$ denotes the Fourier 
transform with respect to time of $h(t)$.  Since $h_+(t)$ and $h_\times(t)$ are real-valued functions 
there are also negative frequency terms given by $\tilde h_{+,\times}(-f) = \tilde h_{+,\times}^*(f)$.

\begin{figure*}
\includegraphics[width=\textwidth]{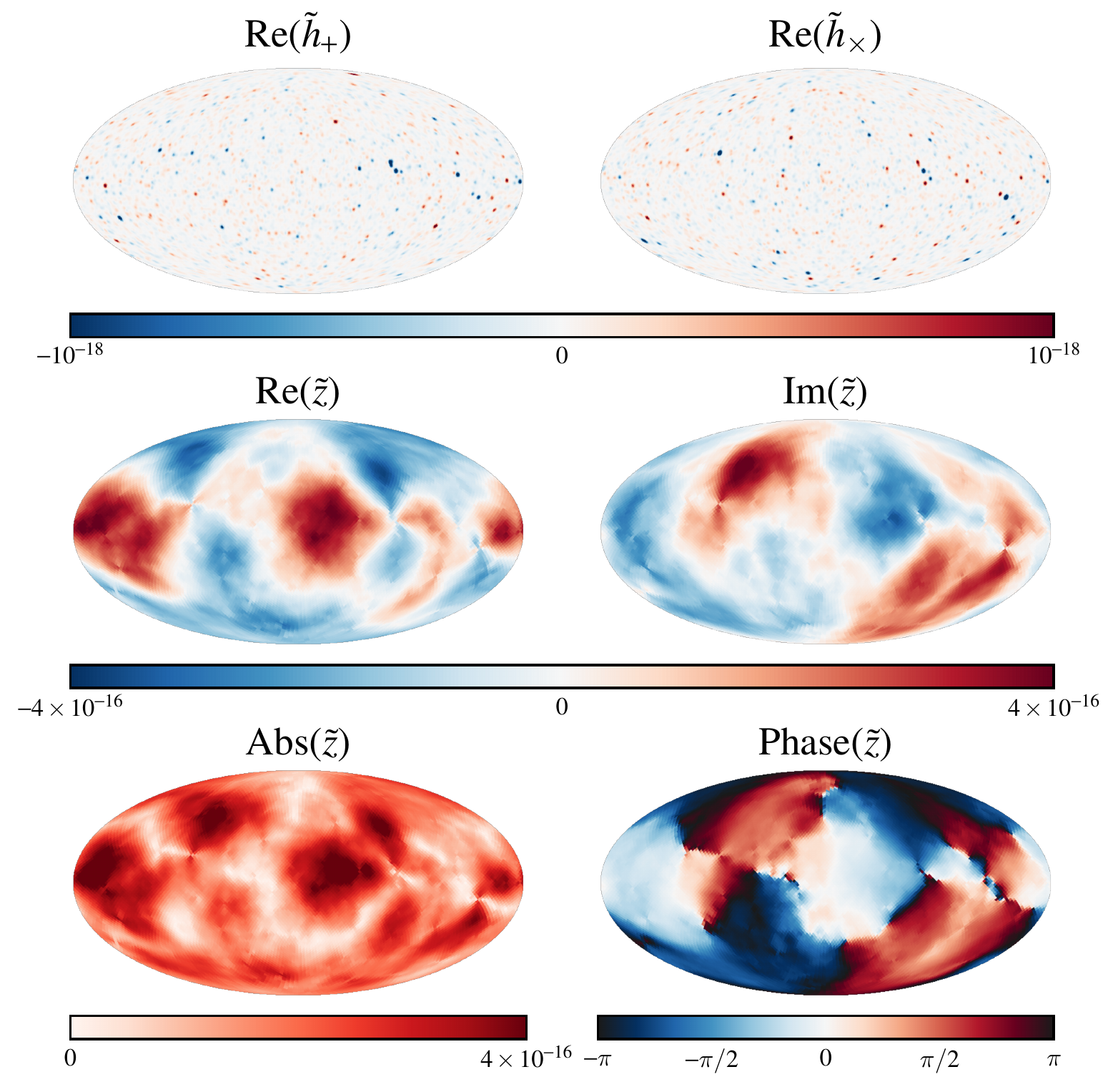}
\caption{A \textsc{gwb} produced by a population of \textsc{smbbh}s in a frequency bin with a central 
	value of $10$~nHz and a width of $1/10$~yr.  The first row shows the real 
	components of $\tilde h_+$ and $\tilde h_\times$ produced at the source locations, smoothed 
	to $2^\circ$ and clipped at an amplitude of $10^{-18}$ to show detail (maxima are $\sim 10^{-16}$).  
	The imaginary components, not pictured,  are similar.  Other rows show the induced redshift map.  
	The middle row shows the real and imaginary components, and the bottom row shows the same 
	map in terms of amplitude and phase. The $\tilde z$ maps show that the background has some
	source confusion, but a handful of the brightest sources contribute most of the signal.  }
\label{fig:pop_gwb}
\end{figure*}

As this gravitational wave (assumed to originate far outside our galaxy) passes a pulsar and the 
earth, the pulse train seen on earth gains a frequency shift of 
\begin{equation}
 \frac{\Delta \nu}{\nu} = z(t, \hat p, \hat k) = \frac{1}{2} \frac{p^i p^j}{1 + \hat k \cdot \hat p} 
	\big[h_{ij}(t_\text{psr}, \hat k) - h_{ij}(t_\text{earth}, \hat k)\big],
\end{equation}
where the direction to the pulsar is written $\hat p$.  Frequency shifts will be of the same order of 
magnitude as $h$, that is $\lesssim 10^{-15}$.  This is too small to measure.  However, 
over many cycles, the frequency shift will affect the time of arrival of the pulses:
\begin{equation}
	r(t, \hat p) = \int_0^t dt' z(t', \hat p),
\end{equation}
producing the \textsc{gw} contribution to \textsc{pta} timing residuals.  The amplitude of $r$ will be 
of order $h/f \lesssim 100$~ns for waves with $f\sim10$~nHz.

%If we assume that gravitational wave sources do not evolve 
%between the time they pass the earth and all pulsars, 
Since the \textsc{gw}s will pass through our entire galaxy,
$z(t, \hat p)$ (and equivalently, $r$) can be split into two terms: the term due to the metric 
disturbance at the earth, $h_{ij}(t_\text{earth}, \hat k)$, and the term at the pulsar.  Earth terms due 
to the same \textsc{gw} will be correlated between different points on the sky (different pulsars), but
pulsar terms will depend on the distance between the earth and the pulsar.  Absent detailed information
about pulsar distances, and assuming that the sources do not evolve significantly in frequency between 
the time that the waves pass the earth and all pulsars,
pulsar terms can be modeled as a term of the same magnitude as the earth 
term but with a random additional phase.  For simplicity, the rest of the paper will concentrate on 
the earth terms, which are correlated on the sky, although they can also be considered as a form 
of self-noise, which would enter the calculations in \autoref{sec:Cl_SN}.  

Assuming circular binaries, we write
\begin{equation}
\tilde z_\text{earth}(f, \hat p, \hat k) = \frac{1}{2} \frac{p^i p^j}{1 + \hat k \cdot \hat p} \tilde h_{ij}(f, \hat k),
\label{eq:z_def}
\end{equation}
where $\tilde h_{ij}(f, \hat k)$ is of the form given in \autoref{eq:h(f)}.  This is a complex scalar field.
Calculating the total redshift induced in any one direction $\hat p$ requires integration over all $h_{ij}$ 
coming from all directions $\hat k$.

We present two examples of $\tilde z_\text{earth}$ in a single frequency bin.  \autoref{fig:two_srcs} 
shows the redshift map produced by two \textsc{smbbh} sources of the same amplitude but different 
inclinations and random other parameters.  \autoref{fig:pop_gwb} is an example of a likely \textsc{gwb}
for a single frequency bin. It is generated from the population models of \citet{roebber16}.  Every binary 
black hole is assigned a random set of parameters. 
Although the population contains 150,000 \textsc{gw} sources with $\mchirp > 10^7 M_\odot$, 
relatively few are visible in the maps.

\section{Harmonic analysis of redshift maps}
\label{sec:Cl_analysis}
We consider two toy model \textsc{gwb}s which can be considered as limiting cases for a 
stochastic \textsc{gwb} produced by a population of \textsc{smbbh}s with no underlying anisotropy.  The first
example is for a single source of \textsc{gw}s, which is an idealization of the case where the \textsc{gw}
power in a frequency bin is dominated by a single bright source.  The second example is the canonical 
case where the \textsc{gwb} is a stochastic Gaussian random field.  This represents the opposite limit of a 
confusion background, which has no visible individual sources.

\subsection{A single gravitational wave source}
For the case of a single source, we will consider a single inspiraling pair of \textsc{smbbh}s 
located at the north pole and aligned with our $(l,m)$ coordinate choices, so that $\psi=0$ and $\phi=0$.  
This choice will allow us to do the calculations in a simple form; all other possible single sources can be 
reproduced by applying a rotation at the end. 

\begin{figure}
\begin{center}
\includegraphics{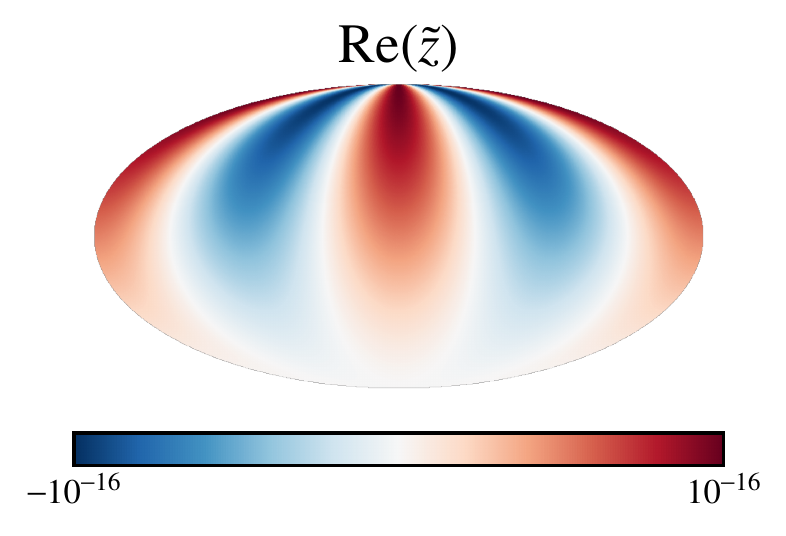}
\end{center}
\caption{The redshift pattern produced for an edge-on single source at the north pole with 
\textsc{gw} amplitude $\mathcal{A}=10^{-16}$ and $\psi=0$. For this example, there is no 
	imaginary component. The integrated redshift term which affects the pulsar timing residuals 
	looks very similar, but has a maximum amplitude of $\mathcal{A}/2\pi i f$, rather than $\mathcal{A}$.
}
\label{fig:single_source}
\end{figure}

In this coordinate system, the direction of the \textsc{gw} propagation is $\hat k = - \hat z$ and the 
vectors defining the polarization are $(l, m) = (-\hat x, \hat y)$.  Plugging these definitions,  
\autoref{eq:e+ex}, and \autoref{eq:hab_def} into \autoref{eq:z_def} and considering the response 
in all directions $\hat r = (\theta, \phi)$ produces the redshift induced across the sky by a single 
source at the north pole:
\begin{equation}
	\tilde z(\theta, \phi) = \frac{1}{2} (1+\cos\theta) (\cos 2\phi \, \tilde h_+ - \sin2\phi \, \tilde h_\times).
	\label{eq:zthetaphi}
\end{equation}
This is a continuous field everywhere except in the direction of the source, where the $\theta$ term is 
constant, but the $\phi$ term is undefined due to rapid oscillation at small $\theta$.  
See \autoref{fig:single_source}.

Since \autoref{eq:zthetaphi} is a scalar field, it can be represented as the sum of spherical harmonics.
Doing this expansion (see \autoref{sec:almcalc}) produces
\begin{equation}
	a_{lm} = 2\pi \sqrt{ \frac{2l + 1}{4\pi} \frac{(l-2)!}{(l+2)!} } (\tilde h_+ \pm i \tilde h_\times), \text{ for } m=\pm2.
	\label{eq:alm1src}
\end{equation}
Note that the subscripts $l, m$ here are the usual spherical harmonic labels and not tensor indices.  
Interestingly, the $a_{lm}$ only exist for $m\pm2$.  This is a reflection of the four stripes seen in the 
half-beachball form of the pulsar response function (see \autoref{fig:single_source}).  
Fundamentally, this is due to the spin-2 nature of 
gravitational waves---a rotation of the $(\hat l, \hat m)$ coordinate system by $180^\circ$ must produce 
the same result. 

Although the underlying form of the redshift pattern is fundamental, 
the representation in spherical harmonics is a result of our choice to place the \textsc{gw} source 
at the north pole.  A source located elsewhere in the sky can be expressed by a rotation of 
\autoref{eq:alm1src}.  This will mix between $m$ components, so that a generic source will require 
a full set of spherical harmonics to reproduce its response function.  However, since rotations of 
spherical harmonics cannot transform one $l$ to another, the scaling of $a_{lm}$ with $l$ will remain 
consistent.

A statistical description of the $l$-scaling of $z$ can be found in the angular power spectrum \citep{dodelson03}:
\begin{equation}
	C_l = \frac{1}{2l + 1} \sum_m a_{lm}^* a_{lm}
\end{equation}
For the case of a single source, this becomes:
\begin{equation}
	C_l = \frac{2\pi \left( \tilde h_+^2 + \tilde h_\times^2 \right)}{(l+2)(l+1)(l)(l-1)}, \text{ for } l\geq 2.
\label{eq:cl}
\end{equation}
This is a steeply decreasing function of $l$.  Since it is only a function of $l$ it does not vary 
under rotations.  Therefore, \autoref{eq:cl} holds for any single \textsc{smbbh} source of 
\textsc{gw}s with polarizations $\tilde h_+$ and $\tilde h_\times$.

\subsection{A statistically isotropic Gaussian random field gravitational wave background}

The second case that we consider is the case where the \textsc{gwb} is a statistically isotropic 
Gaussian random field.  This represents an idealization of a stochastic background produced by 
many sources, and similar models have frequently been considered in the \textsc{pta} literature. 
Our discussion will follow \citet{burke75}.

This kind of background is the most similar to the \textsc{cmb}.  However, a major difference is that the
\textsc{gwb} is stationary but not time-invariant on observational timescales.  (This is because 
it is produced by rotating \textsc{smbbh}s, which cause the background to rotate through polarizations).
When we Fourier transform the data to work with a single frequency bin, the resulting maps will 
be complex, unlike the real \textsc{cmb} maps. 

A \textsc{gwb} produced by \textsc{smbbh}s produces a redshift field equal to the sum over the
redshift field produced by each source.  Every source will produce a redshift field of the form of 
\autoref{eq:zthetaphi}, but with an additional random rotation, which sets $(\theta, \phi, \psi)$ to 
random new values.  For a field with maximal source confusion, we consider independent sources 
of similar amplitude along every line of sight. 

\begin{figure}
	\includegraphics{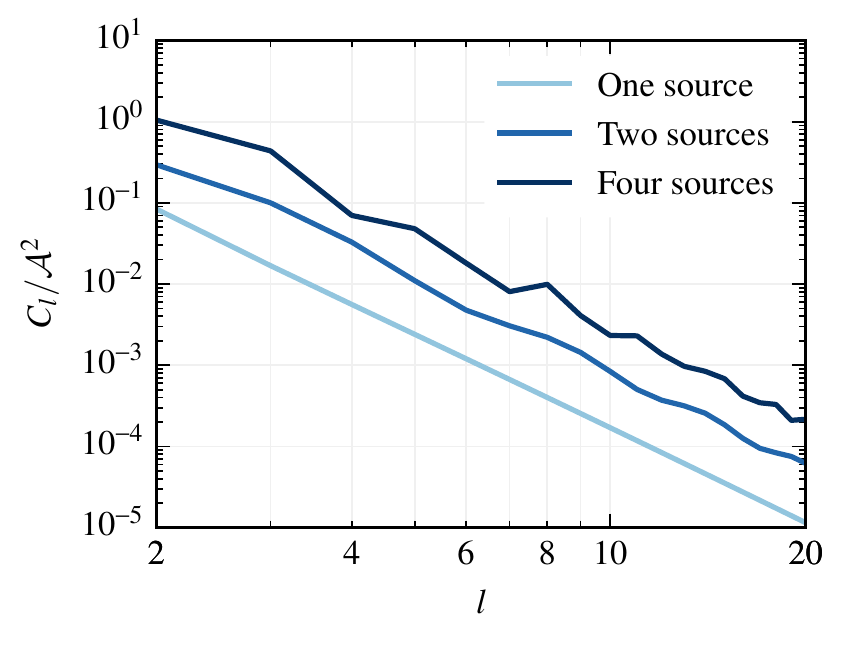}
	\caption{The power spectrum for $z$ maps containing several random sources with 
		amplitude $\mathcal{A}$. One source produces a power spectrum that scales exactly 
		as $[(l+2)(l+1)(l)(l-1)]^{-1}$.  Adding a second source has two effects: increasing the 
		amplitude and inducing small interference ripples around the fiducial power law.  
		Adding more sources to the map will continually increase the amplitude and change 
		the ripples, but they will remain a secondary effect.}
	\label{fig:2src_Cl}
\end{figure}

As we incoherently  add sources, the value of each of the $a_{lm}$'s will change.  However, since 
the $l$-dependence is left unchanged after a rotation, the new $a_{lm}$'s will be given by a sum of 
terms with varying complex amplitude, but which all scale with $l$ in the same way as \autoref{eq:alm1src}.  
In this way we can see that adding sources preserves the shape of the average power spectrum.  
An example is shown in \autoref{fig:2src_Cl}.

The amplitude of the average power spectrum is however strongly affected by the number (and strength) 
of \textsc{gw} sources.  For fixed $l$, adding randomly-located sources can be modeled as a random
walk in the amplitudes of each mode.  The random walk will have a mean of zero, but the variance will 
increase proportionally to the number of sources.  Since the power spectrum is the variance of the 
$a_{lm}$, a Gaussian random field produced by $N_\text{src}$ identical sources should have 
an underlying power spectrum of the form:
\begin{equation}
	C_l \propto \frac{N_\text{src}}{(l+2)(l+1)(l)(l-1)}.
	\label{eq:Cl_gaussian}
\end{equation}

In other words, adding many sources of similar amplitude randomly and incoherently will produce a 
field whose  harmonic decomposition is made up of terms arbitrarily drawn from a distribution given 
by \autoref{eq:Cl_gaussian}.  In real space, this means that as the number of sources increase, points 
separated by given angle will maintain an average correlation, but actual values will be randomly 
distributed according to a Gaussian distribution.

\begin{figure}
	\includegraphics{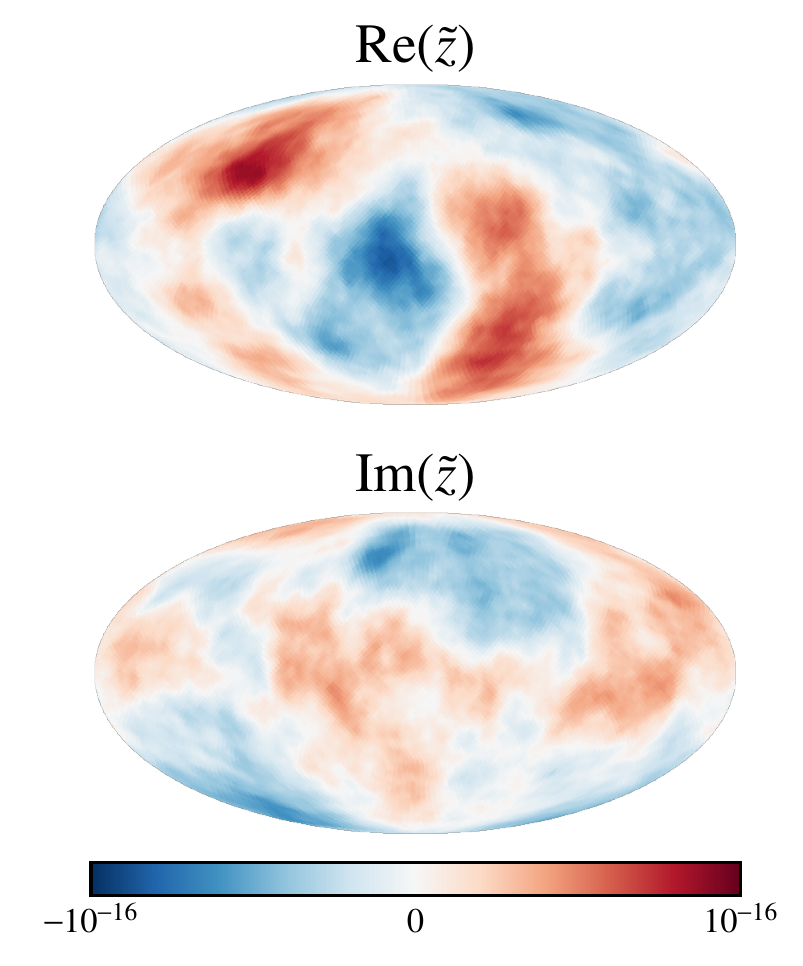}
	\caption{A single realization of a statistically-isotropic gaussian random redshift map
		with a power spectrum given by \autoref{eq:Cl_gaussian}.}
	\label{fig:grf}
\end{figure}

A fully Gaussian random field is shown in  \autoref{fig:grf}.  By comparing \autoref{fig:grf} and 
the low-frequency population model \textsc{gwb} shown in \autoref{fig:pop_gwb}, we see that 
the population produces a mostly-Gaussian field, but with small artifacts around the brightest 
sources.  While the distribution of source amplitudes in a real population is steeply decreasing, 
it is still true that a relatively small number of sources produce a majority of the signal. 

It is important to recall that while \autoref{eq:Cl_gaussian} gives the expectation value for the power 
spectrum of the field, any single realization will only approximately reproduce it.  We will discuss this 
further in \autoref{sec:gaussian_variance}.

\section{Variance in the power spectrum and the Hellings \& Downs curve}
\label{sec:variance_Cl_HD}

In this section we will discuss the relationship between the power spectrum of the redshift 
map and the Hellings \& Downs curve.  We will explore how variance around the 
fiducial power law power spectrum affects the shape of the two-point correlation function 
in the \textsc{gwb} models previously discussed.  Since redshifts may be readily converted 
into timing residuals independently of angle, the following analysis applies to both.  

\subsection{The power spectrum is the harmonic transform of the Hellings \& Downs curve}

In the standard \textsc{pta} analysis, timing residuals from different pulsars are correlated.  This
process should average away effects of noise, which is expected to be uncorrelated between 
different pulsars \citep[e.g.][]{lommen15}.  Correlations due to passing \textsc{gw}s in the timing 
residuals of two pulsars separated by an angle $\theta$ are expected to take the form of 
the Hellings \& Downs curve \citep{hellings83}:
\begin{equation}
C(\theta) = \frac{1}{2}\left\{ 1 + \frac{3}{2}(1-\cos\theta)
			\left[ \ln\left(\frac{1-\cos\theta}{2}\right) - \frac{1}{6}  \right] \right\}.
\end{equation}

This is a real-space two-point correlation function (also sometimes referred to as an overlap reduction function).  
It is an especially useful statistic for a Gaussian
random field, since the statistics of such a field can be described entirely by the mean (one-point function) 
and standard deviation (two-point function),  e.g. \citet{allen99}. For symmetric fields such as the 
\textsc{gwb} or the cosmic microwave background, the mean vanishes, ensuring that the two-point 
function contains the entire statistical information of the field.  

If we consider taking the two-point function of a large number of points across the sky (e.g.\ the 
pixels in a map such as those shown in \autoref{fig:two_srcs}), it becomes sensible to define 
a harmonic-space analog of the two point correlation function.  This function is the power spectrum
discussed earlier, and the conversion between the two forms may be written \citep{dodelson03}:
\begin{equation}
	C(\theta) = \sum_{l=0}^\infty C_l \frac{2l+1}{4\pi} P_l(\cos\theta), 
	\label{eq:2pt_def}
\end{equation}
where $C_l$ is given \autoref{eq:Cl_gaussian}.  

A proof for this relation is shown in \citet{gair14}, although they use different notation: their $C_l$
is a constant since they are concerned with the power spectrum of the \textsc{gw} point source 
distribution. Their additional factor of $N_l^2$ is equivalent to the $l$-scaling in our choice of $C_l$, 
and represents the effect of the pulsar response function.  

An additional concern is the normalization of  \autoref{eq:Cl_gaussian} required to reproduce the 
standard form of the Hellings \& Downs relation. This can be found from \autoref{eq:2pt_def}, 
as done by \citet{gair14}, or by considering Parseval's theorem for spherical harmonics:
\begin{align}
 \sum_{l=0}^\infty \sum_{m=-l}^l |a_{lm}|^2 &= \int_{S^2} d\Omega \, |\tilde z(\theta, \phi)|^2 \nonumber \\
 \sum_{l=2}^\infty (2l + 1) C_l  &= 4\pi \,C(0).
\end{align}
Doing the sum produces a factor of $1/3$, so the normalization of the power spectrum required 
to satisfy this constraint is
\begin{equation}
	C_l =  \frac{6\pi}{(l+2)(l+1)(l)(l-1)}.
	\label{eq:Cl_norm}
\end{equation}

Since the Hellings \& Downs curve is the map space version of the expected form of the 
angular power spectrum, any effects which modify the power spectrum can be converted 
into potentially measurable effects on the two-point correlation function.

\subsection{Variance in the Hellings \& Downs curve for a single gravitational wave source}

For a single source of gravitational waves at the north pole, we were able to calculate the 
$a_{lm}$ exactly (up to a rotation).  The only uncertainty left is in the amplitudes of the two 
polarizations, which will be specified by the value of the parameters 
$\mathcal{A}, \Phi_0, \iota, \psi$ for any single source.  Since there is no uncertainty in the 
underlying $a_{lm}$, the form of the $C_l$ is given by \autoref{eq:cl}, with no variance.  
This will produce a real-space two-point correlation that is \emph{exactly} equivalent to the 
Hellings \& Downs curve.  

If we add a second source to the map, the form of the $a_{lm}$ gains additional degrees of
freedom relating to the angle between the two sources and their relative orientations.  In 
general, the two redshift patterns will interfere, leading to maps like those in \autoref{fig:two_srcs} 
and power spectra similar to \autoref{fig:2src_Cl}.  The primary effect on the power spectrum
is to roughly double its amplitude (depending on parameter values).  The ripples induced
by the interference are typically a smaller effect.  

As a result, the primary effect on the two-point correlation is to increase the signal.  The 
small ripples will result in a mild change in the shape of the correlation function away from
Hellings \& Downs.  This will be discussed in greater detail in the next section.

When the two sources are appropriately aligned, more dramatic effects can be produced.  
Co-located sources with out-of phase redshift patterns can cancel, and sources separated by 
$90^\circ$ can have power spectrum oscillations of $20$--$50\%$ at low $l$.  In these cases, 
the two-point correlation will either have decreased signal (as in the first case) or the shape 
will change significantly (the second case).

\subsection{Variance in the Hellings \& Downs curve for a Gaussian random field}
\label{sec:gaussian_variance}

For a statistically isotropic Gaussian random gravitational wave field, the expectation value 
of the power spectrum will be of the form of \autoref{eq:Cl_gaussian}.
However, we are able to observe only one realization of the \textsc{gwb}.  

For a Gaussian random background, all multipole moments $a_{lm}$ are drawn from a 
Gaussian distribution with variance $C_l$.  Even if we are able to measure the $a_{lm}$
perfectly, our ability to correctly estimate the variance of the distribution will be affected
by the number of modes for each value of $l$.  Therefore, the observed power 
spectrum $\hat C_l$ will not be of the same form as the expectation value $C_l$.  

In contrast, for a single source, the choice of each $a_{lm}$ depends on the sky 
location and polarization angle of the source, but is otherwise entirely set.  The distribution
is entirely random for a Gaussian field, and entirely non-random for a single source.

This limitation on the measurability of $C_l$ is the cosmic variance familiar from 
calculations of the cosmic microwave background power spectrum: 
\begin{equation}
	\Delta C_l =  \frac{C_l}{\sqrt{2l+1} }.
	\label{eq:cv_def}
\end{equation}
This equation differs from the standard definition \citep[e.g.][]{dodelson03} by a factor of 
$\sqrt{2}$.  Since the Fourier-transformed \textsc{gwb} is a complex field, it contains twice 
the information of a real field such as the \textsc{cmb}.

The cosmic variance represents the range within which an observed $\hat C_l$ is expected 
to differ from the true unobservable $C_l$ for each $l$.  When $\hat C_l$ no longer follows 
\autoref{eq:Cl_gaussian}, $\hat C(\theta)$ will no longer follow the Hellings \& Downs curve.
By changing a single multipole, we are effectively changing the weights of individual 
Legendre polynomials in \autoref{eq:2pt_def}.

\begin{figure}
\begin{center}
\includegraphics{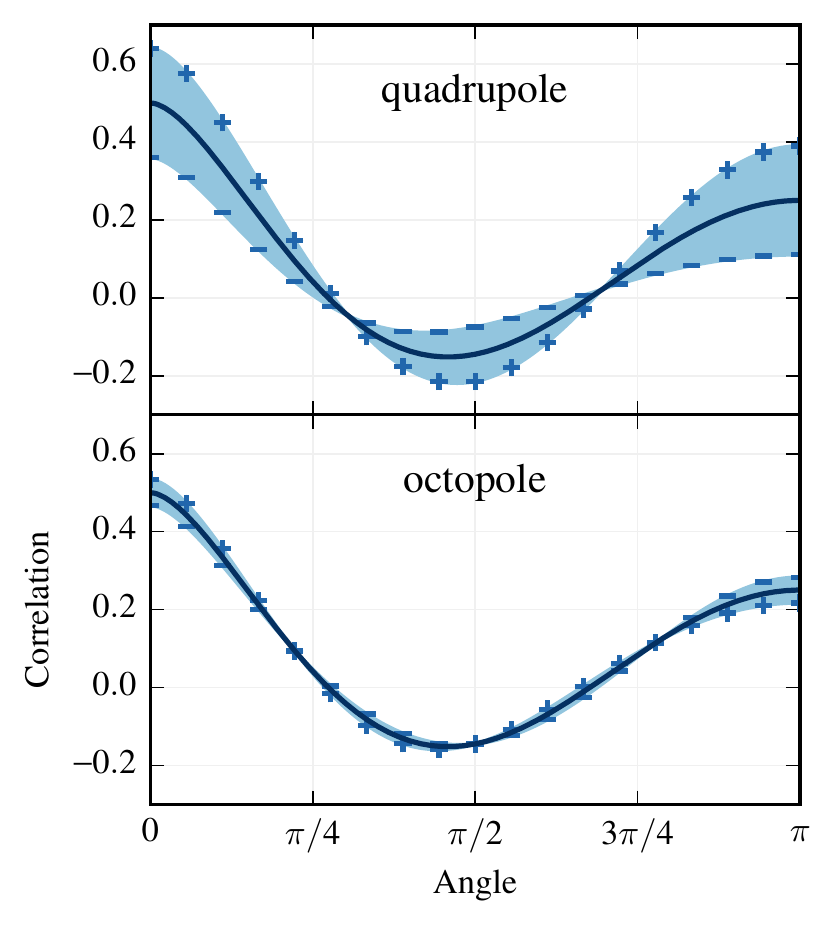}
\end{center}
\caption{The effect on the Hellings \& Downs curve due to changing a single term of the power 
	spectrum by an amount within the cosmic variance.  The solid dark line is the Hellings \&  
	Downs curve and the edges of the shaded region marked with $+$ or $-$ respectively 
	represent the effects of increasing or decreasing the power spectrum term.}
\label{fig:hd_cv}
\end{figure}

Since $C_l$, and consequently $\Delta C_l$, is a strong function of $l$, the effect of cosmic
variance will be strongest for the first few multipoles.  This is shown in \autoref{fig:hd_cv}. 
The quadrupole term is by far the most important, but combinations of several other terms can 
also affect the shape of the two-point correlation function.

It will often be the case that the power spectrum of a Gaussian random \textsc{gwb} will 
have a low or high quadrupole or octopole by chance.  It would therefore not be surprising
to have a two-point correlation function which does not match the Hellings \& Downs curve.
Note that in contrast to the work in \citet{mingarelli13,taylor13, gair14}, this change in shape 
of the two-point correlation function is not due to large-scale anisotropy in the source population, 
but occurs even in statistically-isotropic \textsc{gwb}s.

For two sources which induce a noticeable shape shift on the two point correlation function, the
basic mechanism is the same as for the Gaussian case: specific Legendre polynomials in the 
expansion of \autoref{eq:2pt_def} are being up- or downweighted.  The primary difference in 
these two cases is that for a Gaussian field, the amount by which each $C_l$ varies from the 
expectation value is independent of all the others.  For two sources, the specific interference 
pattern between the two redshift maps leads to oscillations in the power spectrum.  These 
oscillations are set by the relative orientation and distance of the sources and are not random
and not independent. 

So far we have been discussing a \textsc{gwb} composed of a single frequency bin.  However,
\textsc{pta}s are typically sensitive to a range of frequencies.  If all the frequency bins under 
consideration can be described by the same underlying distribution, the effect of cosmic variance
can be ameliorated.  This is because the separate frequency bins can be considered as 
independent realizations of the same map---including more frequency bins allows us to sample the 
distributions more accurately.  From \autoref{eq:cv_def}, $\Delta C_l \sim (N_\text{modes})^{-1/2},$ 
so using $n$ similar bins in the analysis will reduce the cosmic variance by a factor of $\sqrt{n}$.  
Even for a single frequency bin, an analysis assuming Hellings \& Downs behavior may be 
sufficient to allow an initial detection \citep{cornish16}.

\section{How well can a \textsc{pta} measure the angular power spectrum?}
\label{sec:Cl_SN}

Our analysis has focused on the analysis of redshift maps in harmonic space, inspired by 
\textsc{cmb} analyses.  However, unlike \textsc{cmb} experiments which make measurements 
over large regions of the sky, \textsc{pta}s are only sensitive to the redshift field in the direction
of its pulsars.  The observed field is a partial sky map, sampled at $M$ discrete sky locations 
corresponding to the positions of the pulsars in the array.  

The number of pulsars will limit the degree to which a \textsc{pta} can measure harmonics 
of the \textsc{gwb}, but the steepness of the power spectrum will turn out to be a more 
important limitation.  We estimate \textsc{pta} sensitivity to the power spectrum through the 
following signal-to-noise calculation.

For a sparse sampling of the sky, an estimate of a spherical harmonic expansion of
a field $r(\theta, \phi)$ sampled at points $(\theta_i, \phi_i)$ can
be constructed as
\begin{equation}
\hat{a}_{lm} = \frac{4\pi}{\sum_i w_i} \sum_i w_i Y_{lm}(\theta_i, \phi_i) r(\theta_i,\phi_i), 
\end{equation}
where $w_i$ are weights that can be tuned for each point. The minimum variance estimate
will have $w_i$ equal to $1/\sigma_i^2$, where $\sigma_i$ is the variance at each point.
The formally optimal solution would have $\sigma_i$ only including detector noise and terms
intrinsic to the pulsar. However, in practice it would be difficult to separate a given
pulsar's noise properties from a gravitational wave background. 

For simplicity, we assume that all pulsars have equal weight, with rms noise of each pulsar
(for gravitational waves plus noise) $\sigma_0$. Generalizing to varying noise levels
is straightforward.
For $M$ pulsars, the estimated angular power spectrum will then have a noise bias:
\begin{equation}
C_l^N = \frac{4 \pi}{M} \sigma_0^2  \ .
\end{equation}

We can use \autoref{eq:cv_def} to estimate the signal to noise of the amplitude 
for a \textsc{pta}, taking
care to realize that the relevant $C_\ell$ for the noise estimate 
is the combined signal and noise power spectrum.
For the signal power spectrum, we know that the form should follow from \autoref{eq:Cl_norm}.  
Given a variance in residuals from gravitational waves $\sigma_{gw}^2$, we write 
\begin{equation}
C_l^S = \sigma_{gw}^2 C_l.
\end{equation}

The resulting estimate for the signal
to noise for a given multipole $l$ is
\begin{equation}
\left(\frac{S}{N}\right)_l^2 = (C_l^S)^2 \frac{2l+1}{(C_l^S+C_l^N)^2}
\end{equation}

For a first detection, the expectation is that the noise power in the large-scale 
correlated timing residuals will be much larger than the signal power. We can then
simplify the signal-to-noise estimate by dropping the signal part of the last term.
Explicitly, the signal-to-noise in the limit of a weak detection is
\begin{equation}
\left(\frac{S}{N}\right)_{l; \text{ weak}}^2 = \frac{9}{4} \frac{2l+1}{[(l-1)l(l+1)(l+2)]^2} \frac{M^2 \sigma_{gw}^4}{\sigma_0^4} 
\end{equation}
Summing this over all $\ell$ gives a numerical prefactor of $1/48$, with the $l=2$
term alone contributing $5/256$. The $l=2$ term
thus contributes 93.75\% of the $(S/N)^2$. If one only measured the power in the quadrupole
anisotropy of the timing residuals, the resulting signal-to-noise would
be 97\% of the total signal-to-noise available. This is simply because the quadrupole is
contributing such a large fraction of the total power that it is far and away the largest
signal to be measured and the signal-to-noise adds in quadrature rather than linearly.

To compare with previous work, we can do the similar calculation in map space. 
As shown in \citet{siemens13}, the comparable prefactor for this calculation in map space reduces
to the total number of pairs times the mean of the square of the Hellings \& Downs curve.
For a full-sky survey, the mean of the square of the Hellings \& Downs curve is 1/48, 
while the number of unique pulsar pairs is $M(M-1)/2$, very close to $M^2/2$, with the
$(M-1)$ instead of $M$ coming from the explicit nulling of autocorrelations in the calculation.

\section{Discussion}
\label{sec:discussion}

In this work we have introduced an alternate framework for considering spatial variation in 
gravitational wave backgrounds.  We primarily work with the all-sky redshift patterns induced
by gravitational waves passing the earth.  Using standard techniques from \textsc{cmb} analysis, 
we do all calculations in harmonic space for computational simplicity, but convert to map space 
to discuss measurable quantities.  Since we assume non-evolving \textsc{gw} sources, all
results are also true for the earth term of the expected pulsar timing residuals, up to a normalization.  
This assumption breaks down for rare high-mass, high-frequency binaries which evolve on 
timescales of $\sim$~kyr rather than $\sim$~Myr \citep{mingarelli12}.
 
We explicitly decomposed the redshift pattern produced by a single source of \textsc{gw}s 
into spherical harmonics, which allowed us to calculate the power spectrum of a single source's 
redshift map exactly.  We showed that the expectation value of the power spectrum for a statistically 
isotropic gaussian random \textsc{gwb} has the same form as for a single source.  Using the relation 
between the power spectrum and the real space two-point correlation function, we explored the 
degree to which variance in the power spectrum changes the shape of the two-point correlation 
function away from Hellings \& Downs.  In particular, cosmic variance in the quadrupole moment of 
the power spectrum for a Gaussian random field can have significant effects on the amplitude of the
curve, while also changing its shape.  Finally, we showed that the quadrupole term of the power 
spectrum contributes $97\%$ of the signal-to-noise measured by a \textsc{pta}.

Throughout this work, we have treated the \textsc{gwb} as one of two idealized cases: a single source
or a Gaussian random field.  A \textsc{gwb} produced by a population of sources will lie somewhere 
between these two cases, as suggested by \autoref{fig:pop_gwb}.  It is likely that the degree to 
which a population of sources resembles one case or another changes as a function of frequency, 
with shot noise in the \textsc{smbbh} population becoming more important at higher frequencies.

We have confirmed that \textsc{gwb}s dominated by a single bright source, which are highly anisotropic 
and non-Gaussian, and those which are isotropic, unpolarized, and Gaussian look very similar from the 
point of view of a two-point correlation function, as previously reported by \citet{cornish13}.  This 
suggests that two-point correlation functions will be effective for detecting \textsc{gwb}s of all kinds.
But they will be ineffective for characterizing \textsc{gwb}s and searching for single sources, despite
the clear visual differences between \autoref{fig:single_source} and \autoref{fig:grf}. 

This difference should be measurable given a sufficiently high significance measurement of the 
\textsc{gwb} and some luck in its orientation with respect to low-noise pulsars.
A particularly clear example is given in \citet{boyle12}: consider the timing residuals for four 
pulsars, each of which is located in a different stripe near the top of the map in \autoref{fig:single_source}.
The \textsc{gw} signal in each pulsar will be perfectly correlated, differing only by a phase factor 
of $180^\circ$ between adjacent stripes.  No such perfect (anti-)correlation is possible for nearby 
pulsars affected by a Gaussian field such as in \autoref{fig:grf}.

An important difference between these two types of \textsc{gwb}s is that the redshift map produced
by a single source is highly nongaussian.  Although symmetric Gaussian distributions can be 
statistically completely described by their two-point functions, non-Gaussian distributions may 
have higher moments.  Indeed, the example given by \citet{boyle12} is a kind of four-point function.
Future work will explore higher-order correlation functions as a means of characterizing the
degree to which a \textsc{gwb} has Gaussian or point-source-like characteristics.

\acknowledgements{
We thank Vicky Kaspi, Chiara Mingarelli, and Pat Scott for useful discussions and comments.
We acknowledge the support of NSERC and the 
Canadian Institute for Advanced Research
and resources provided by Calcul Qu\'ebec. 
}

\software{healpy \citep{gorski05}, 
matplotlib \citep{matplotlib}, 
numpy, pandas.}

\appendix
\section{Calculating the harmonic expansion of \lowercase{$\tilde z(\theta, \phi)$} for a single source GWB}
\label{sec:almcalc}

From \autoref{eq:zthetaphi}, we have the redshift induced in a direction $(\theta, \phi)$
by a source located at the north pole.  This is a complex scalar field, and can be 
expanded in spherical harmonics with coefficients:
\begin{align}
a_{lm} & = \int d\Omega \, \left[ \frac{1}{2}(1+\cos\theta)(\cos 2\phi \, h_+ - \sin 2\phi \, h_\times) \right] 
		Y_{lm}^* \\
	& = \underbrace{\frac{1}{2} (-1)^m \sqrt{ \frac{2l+1}{4\pi} \frac{(l-m)!}{(l+m)!} } }_\text{A} \;
		\overbrace{\int_0^{2\pi} d\phi \, e^{-im\phi} 
			\left( \cos 2\phi \, \tilde h_+ - \sin 2\phi \, \tilde h_\times \right)}^\text{B} \;
		\underbrace{\int_0^\pi d\theta \, \sin\theta \, (1+\cos\theta)P_l^m(\cos\theta)}_\text{C},
\end{align}
where $P_l^m(\cos\theta)$ are the associated Legendre polynomials.  This factorizes into two 
integrals ($B$, $C$) and one constant term ($A$).  
Beginning with the integral over $\phi$, we find that it simplifies to 
\begin{align}
B = \frac{1}{2}\left(\tilde h_+ + i \tilde h_\times\right)  \int_0^{2\pi} d\phi \, e^{-i\phi(m-2)} +  
	\frac{1}{2}\left(\tilde h_+ - i \tilde h_\times\right)  \int_0^{2\pi} d\phi \, e^{-i\phi(m+2)}.
\end{align}
Since $\int_0^{2\pi} d\phi \, \exp(-\alpha i\phi) = 2\pi \,\delta(\alpha)$ for 
real $\alpha$, only terms with $m=\pm 2$ exist. They
are given by
\begin{equation}
	B = 
	\begin{cases}
		\pi(\tilde h_+ \pm i\tilde h_\times),	& m = \pm 2 \\
		0,					& \text{all other $m$}
	\end{cases}.
\end{equation}

This constraint on $m$ allows us to simplify both our constant term and $\theta$ integral:
\begin{align}
A\cdot C 	& = \frac{1}{2} \sqrt{ \frac{2l+1}{4\pi} \frac{(l\mp2)!}{(l\pm2)!} }
			\int_0^\pi d\theta \, \sin\theta (1+\cos\theta)P_l^{\pm2}(\cos\theta) \\
		& = \frac{1}{2} \sqrt{ \frac{2l+1}{4\pi} \frac{(l-2)!}{(l+2)!} }
			\int_0^\pi d\theta \, \sin\theta (1+\cos\theta)P_l^{2}(\cos\theta), \\
\end{align}
using the following property of associated Legendre polynomials:
\begin{equation}
	P_l^{-m} = (-1)^m \frac{(l-m)!}{(l+m)!} P_l^m.
\end{equation}
Writing  $\mu=\cos\theta$, we solve the integral over $\theta$:
\begin{align}
	C 	& = \int_{-1}^1 d\mu \,  (1+\mu)P_l^{2}(\mu) \\
		& = \int_{-1}^1 d\mu \,  (1+\mu) (1-\mu^2) \frac{d^2}{d\mu^2}P_l(\mu)
			& \text{(in terms of ordinary Legendre polynomials)} \\
		& = \int_{-1}^1 d\mu \,  (1+\mu) \left[ 2\mu \frac{d}{d\mu}P_l(\mu) 
			- l(l+1) P_l(\mu) \right] & \text{(using the defining differential equation)} \\
		& = \int_{-1}^1 d\mu \,  (1+\mu) 2\mu \frac{d}{d\mu}P_l(\mu)  
			& \text{(second term zero by orthogonality)} \\
		& = \left[ 2\mu^2P_l(\mu) + 2\mu P_l(\mu) \right]_{-1}^1 -
			\int_{-1}^1 d\mu \, (4\mu + 2) P_l(\mu) \\
		& = 2 \left[ P_l(1) + P_l(1) + P_l(-1) - P_l(-1) \right] & \text{(second term zero by orthogonality)} \\
		&   = 4
\end{align}

Putting everything together, 
\begin{equation}
	a_{lm} = A \cdot B \cdot C = 
	\begin{dcases}
		2\pi \sqrt{ \frac{2l+1}{4\pi} \frac{(l-2)!}{(l+2)!} }
			\big(\tilde h_+ \pm i\tilde h_\times\big), & m = \pm 2, \, l\geq 2. \\
		0,	& \text{otherwise.}
	\end{dcases}
\end{equation}

\bibliography{pta_lm_response}

\end{document}